\documentclass[aps,pre,showpacs,showkeys,twocolumn,superscriptaddress,floats,floatfix]{revtex4-2}

\usepackage{graphicx}
\usepackage{float}
\usepackage{dcolumn}
\usepackage{bm}
\usepackage{subcaption}
\usepackage{caption}
\usepackage{comment}
\usepackage{soul}
\usepackage{ragged2e}
\usepackage{blindtext}
\usepackage{hyperref}
\usepackage{ulem}
\usepackage{amsmath}
\usepackage[dvipsnames]{xcolor}
\usepackage[percent]{overpic}
\def \ii {{\rm i}}
\def \jj {{\rm j}}
\def \uu {{\bf u}}
\def \rr {{\bf r}}
\def \vp {{\rm v_p}}
\def \phat {\hat{\bf p}}
\def \Fij {{\bf F}_{\rm ij}}
\def \etaT {{\boldsymbol \eta}^{\rm T}}
\def \etaR {\eta^{\rm R}}

\begin{document}

\title{Flow-Induced Phase Separation for Active Brownian Particles in Four-Roll-Mill Flow}
\author{Soni D. Prajapati}
\email[Correspondence: ]{sonipdot@gmail.com}
 \affiliation{Department of Physics, Indian Institute of Technology Hyderabad, Hyderabad 502284, India}
 
\author{Kusum Seervi}
\email[Correspondence: ]{kusumseervi5@gmail.com}
 \affiliation{Department of Physics, Indian Institute of Technology Hyderabad, Hyderabad 502284, India}
 
\author{Akshay Bhatnagar}%
\email[Correspondence: ]{akshayphy@gmail.com}
\affiliation{Department of Physics, Indian Institute of Technology Palakkad,
Palakkad 678623, Kerala, India.}

\author{Anupam Gupta}
\email[Correspondence: ]{agupta@phy.iith.ac.in}
\affiliation{Department of Physics, Indian Institute of Technology Hyderabad, Hyderabad 502284, India}
\begin{abstract}
We investigate the collective dynamics of active Brownian particles (ABPs) subjected to a steady two-dimensional four-roll-mill flow using numerical simulations. By varying the packing fraction ($\phi$), we uncover a novel flow-induced phase separation (FIPS) that emerges beyond a critical density ($\rm \phi_c \gtrsim 0.5$). The mean-square displacement (MSD) exhibits an intermediate plateau between ballistic and diffusive regimes, indicating transient trapping and flow-guided clustering. The effective diffusivity decreases quadratically with $\phi$, while the drift velocity remains nearly constant, demonstrating that large-scale transport is primarily dictated by the background flow. Number fluctuations show a crossover from normal to giant scaling, signaling the onset of long-range density inhomogeneities in the FIPS regime. Our findings provide new insights into the coupling between activity, crowding, and flow, offering a unified framework for understanding phase behavior in driven active matter systems.
\end{abstract}


\maketitle

\section{Introduction}
\label{sec:level1}

The study of self-propelled particle systems gained significant attention following the pioneering work of Vicsek et al. \cite{vicsek1995novel}, which introduced a minimal model exhibiting collective motion through local alignment interactions. This line of research revealed that the motion of active particles is not only governed by their
internal propulsion mechanisms, but is also strongly influenced by the
surrounding environment, including background flow fields, confinement, or
hydrodynamic interactions~\cite{toner1995long, toner2005hydrodynamics,
ramaswamy2010mechanics, marchetti2013hydrodynamics}. With the rapid advancement
of computational resources in recent years, active matter has become a thriving
area of research offering deep insights into nonequilibrium statistical
mechanics. 

Active matter refers to systems composed of units that consume energy to
generate motion, keeping the system far from equilibrium
~\cite{vrugt2025exactly, marchetti2013hydrodynamics, bechinger2016active, ramaswamy2017active}.
These systems span an extraordinary range of length
scales\cite{volpe2022active}, starting from synthetic colloidal Janus particles and
catalytic swimmers at the microscale~\cite{howse2007self, palacci2013living}, to
macroscopic systems such as bird flocks, fish schools, and
crowds~\cite{vicsek2012collective, cavagna2014bird}. A unifying feature among these active matter systems is their persistent motion, characterized by a finite
persistence time or length over which particles tend to maintain their
direction~\cite{Fily2012,redner2013reentrant, cates2015motility}.

A special phenomenon in active systems is motility-induced phase
separation (MIPS), in which purely repulsive active particles spontaneously
separate into dense and dilute phases
\cite{Fily2012,redner2013structure,buttinoni2013dynamical}. This occurs when
particles are trapped in dense regions due to a feedback mechanism between
motility and local crowding, a process closely tied to the persistence of their
motion~\cite{palacci2013living, buttinoni2013dynamical, cates2015motility, speck2014effective}.

While the active Brownian particle (ABP) model neglects hydrodynamic, it has
proven to be an indispensable minimal model that captures many features of
active matter \cite{Fily2012,bechinger2016active,speck2014effective}. In ABPs,
each particle self-propels at a constant speed and undergoes rotational
diffusion, breaking detailed balance, and giving rise to non-equilibrium behavior
even in simple settings \cite{Fily2012, prajapati2025effect}.
In real-world scenarios, active particles are not isolated but are often immersed in
complex fluid environments or interact with background flows, due to external
fields, confinement ~\cite{zottl2016emergent, lushi2014fluid, saintillan2018rheology}. The interaction between active particles and background flows can give rise to complex dynamical behaviors that are not captured only by considering the ABP model in the absence of external flow influences.

In our recent work ~\cite{prajapati2025effect}, we studied ABPs in a two-dimensional steady flow field and discovered a new non-equilibrium phase, which we term flow-induced phase separation (FIPS), driven by the ratio of persistence length of ABPs and background flow.
In this study, we investigate the onset and characteristics of the FIPS
phase by systematically varying the packing fraction of
ABPs in a two-dimensional background flow \cite{prajapati2025effect}. As MIPS is
known to be strongly dependent on packing fraction \cite{Fily2012,
redner2013structure, siebert2018critical}, we explore FIPS by varying the packing fractions, which range from $0.1$ to $0.9$ to understand how packing fraction influences clustering and phase
behavior under flow influence.
To characterize the system's behavior across this range of packing fraction, we compute several key
observables: number fluctuations\cite{Fily2012, henkes2011active,
kuroda2023anomalous, Narayan2007}, mean squared displacement (MSD) \cite{prajapati2025effect, Fily2012, howse2007self}, overlap function $\rm Q(t)$ \cite{mandal2020extreme, li2015decoupling, karmakar2011exposingstaticscaleglass}, cluster size distributions (CSD) ~\cite{dolai2018phase, prajapati2025effect}, and Probability distribution function (PDF) of Okubo-Weiss parameter ($\Lambda$)~\cite{picardo2018preferential, picardo2020dynamics}.
\section{Model and Simulations}
\label{section2}

We consider a collection of $\rm N$ interacting ABPs immersed in a two-dimensional incompressible fluid. Each ABP is modeled as a soft disk of diameter $\rm 2a$ that self-propels at a speed $\vp$ along an
 orientation vector $\rm \phat_\ii = (\cos\theta_\ii,
\sin\theta_\ii)$, where the subscript $\rm i = 1,
\dots, N$ denotes the particle index, and $\theta_\ii$ denotes the angle between the particle's orientation and the $x$-axis. To describe the dynamics of ABPs, we use an overdamped Langevin equation with an imposed background flow $\uu(\rr_\ii, t)$. The evolution equation of an ABP with position $\rm \rr_\ii$ and orientation $\rm \theta_\ii$ is given by:
\begin{subequations}
\begin{align}
\rm \dot{\rr}_\ii &= \vp \phat_\ii + \mu_{t} \sum_{\jj \ne \ii} \Fij + \etaT_{\ii}(t) + \uu(\rr_{\ii},t), \label{position_tem} \\
\rm \dot{\theta}_\ii &= \etaR_\ii(t) + \frac{1}{2} \omega(\rr_\ii,t), \label{theta_tem}
\end{align}
\end{subequations}

where $\mu_t$ is the motility. $\omega(\rr_\ii, t) = \nabla \times \uu(\rr, t)$ is the flow vorticity at the position of the particle $\rm \rr_\ii$.
The interaction force $\rm \Fij$ on $\rm i^{th}$ particle due to $\rm j^{th}$ particle
is modeled as a short-range soft repulsion given as:
\begin{equation}
\Fij = 
\begin{cases}
\rm -k (2a - |\rr_{ij}|)\, \hat{\rr}_{ij}, & \text{if } |\rm \rr_{ij}| < 2a, \\
\rm 0, & \text{otherwise},
\end{cases}
\end{equation}
where $\rm k$ is the stiffness of the repulsion, the interparticle separation between
$\rm i^{th}$ and $\rm j^{th}$ particles is $\rm \rr_{ij} = \rr_\jj - \rr_\ii$,
and unit vector is $\rm \hat{\rr}_{ij} = \frac{\rr_{ij}}{|\rr_{ij}|}$.

$\rm \etaT_{\ii}(t)$ and $\rm \etaR_\ii(t)$ are translational and rotational noise, respectively. We focus on the athermal case, where $\rm \etaT_{\ii}(t) = 0$, so only the
rotational noise is considered. $\rm \etaR_\ii(t)$ is modeled as Gaussian white noise with zero mean.
\begin{equation}\label{Rotational_noise_corr}
\langle \etaR_{\ii}(t) \etaR_{\jj}(t') \rangle = 2 \nu_{r} \delta_{\ii \jj} \delta(t-t'),
\end{equation}
where $\nu_r$ is the rotational diffusivity \cite{Fily2012, bechinger2016active, speck2014effective, prajapati2025effect}.

To explore the role of background flow, we consider a 2D four-roll mill flow, which is an analytical solution of the Navier-Stokes equation in the high-viscosity limit~\cite{prajapati2025effect,torney2007transport}. The velocity field of the imposed  four-roll mill flow is given by:
\begin{equation}\label{velocity_field}
\rm (u_x, u_y) = U_0 ( -\sin (x) \cos( y),\ \cos( x) \sin( y) ).
\end{equation}

This flow field yields a periodic vorticity pattern in the square simulation domain of size $\rm [0 ~L]^2$, with $\rm L=2\pi$, consisting of a pair of clockwise and counterclockwise vortices, separated by
narrow regions dominated by
strain~\cite{prajapati2025effect,torney2007transport}. This flow has been
investigated both experimentally and theoretically in the context of passive
tracer mixing~\cite{solomon1988chaotic,solomon2003uniform}. We have set the amplitude of the velocity field to $\rm U_0 = 0.5$.

We use the Euler–Maruyama method to numerically integrate equations (\ref{position_tem}) and (\ref{theta_tem}). The parameters are fixed as $\rm \mu_t = 1$, $\rm k = 100$, and $\rm v_p = 0.2$. Based on the characteristic velocity scales of the ABPs and the background flow, we define a dimensionless scaled speed ${\rm V = \frac{v_p}{U_{rms}}}$, where $\rm U_{rms} = U_0/\sqrt{2}$ ~\cite{prajapati2025effect}. A characteristic timescale for ABPs is given by $\rm \nu_r^{-1}$, while that for the background flow can be defined from the characteristic vorticity $\rm \omega_{rms} = U_0$. The dimensionless scaled time is then defined as the ratio of these two timescales, $\rm \tau = \omega_{rms}/(2\pi \nu_r)$. In all simulations, we set ${\rm V \sim 0.5}$ and $\tau \sim 25$. Based on our experience from prior work, we have chosen this particular $V$ and $\tau$ such that for $\phi=0.7$, we will get a flow-induced phase separation (FIPS), where ABPs try to phase separate in the strain-dominated region of the flow~\cite{prajapati2025effect}. We neglect the back-reaction of particles on the background flow, i.e., the flow field remains unaffected by the presence of ABPs~\cite{prajapati2025effect}. The particle radius is fixed at $\rm a = 0.02$ (except in Fig.~\ref{fig:std_dev}(a) and (b) insets). The packing fraction is defined as $\rm \phi = \frac{N \pi a^2}{L^2}$ and is varied from $0.1$ to $0.8$ by changing the number of particles $\rm N$ (e.g., $\rm N = 21991$ for $\phi = 0.7$). All simulations were performed for $10^4$ time units, which is sufficient to fully capture the diffusive regime. In all simulations, $\phi$ serves as the control parameter with $\phi \in [0.1, 0.9]$. 

\section{Results}
\label{sec:results}

We first validate our model by reproducing the well-established results of athermal phase separation in the absence of background flow. Consistent with previous simulations of active Brownian particles (ABPs), we find that increasing the packing fraction drives a transition from a homogeneous state to a motility-induced phase-separated (MIPS) state~\cite{Fily2012}. As a next step, we introduce these ABPs into a four roll-mill flow. Guided by our earlier work, we choose the parameters $\rm V$ and $\tau$ such that, for $\phi = 0.7$, the system exhibits flow-induced phase separation (FIPS), where ABPs preferentially accumulate in the strain-dominated regions of the flow~\cite{prajapati2025effect}.

In the following subsections, we first present a qualitative comparison between MIPS and FIPS across different values of $\rm \phi$. We then analyze the dynamical behavior of the system using the mean-square displacement (MSD) to estimate the drift velocity and effective diffusivity. The relaxation dynamics are further examined through the overlap function. To quantify structural heterogeneity, we analyze number fluctuations and cluster-size distributions. Finally, to characterize the influence of flow topology on FIPS, we examine the distribution of the Okubo–Weiss parameter.

\begin{figure*}[!htb]
    \centering
    \includegraphics[width=.3\linewidth,height=.3\linewidth]{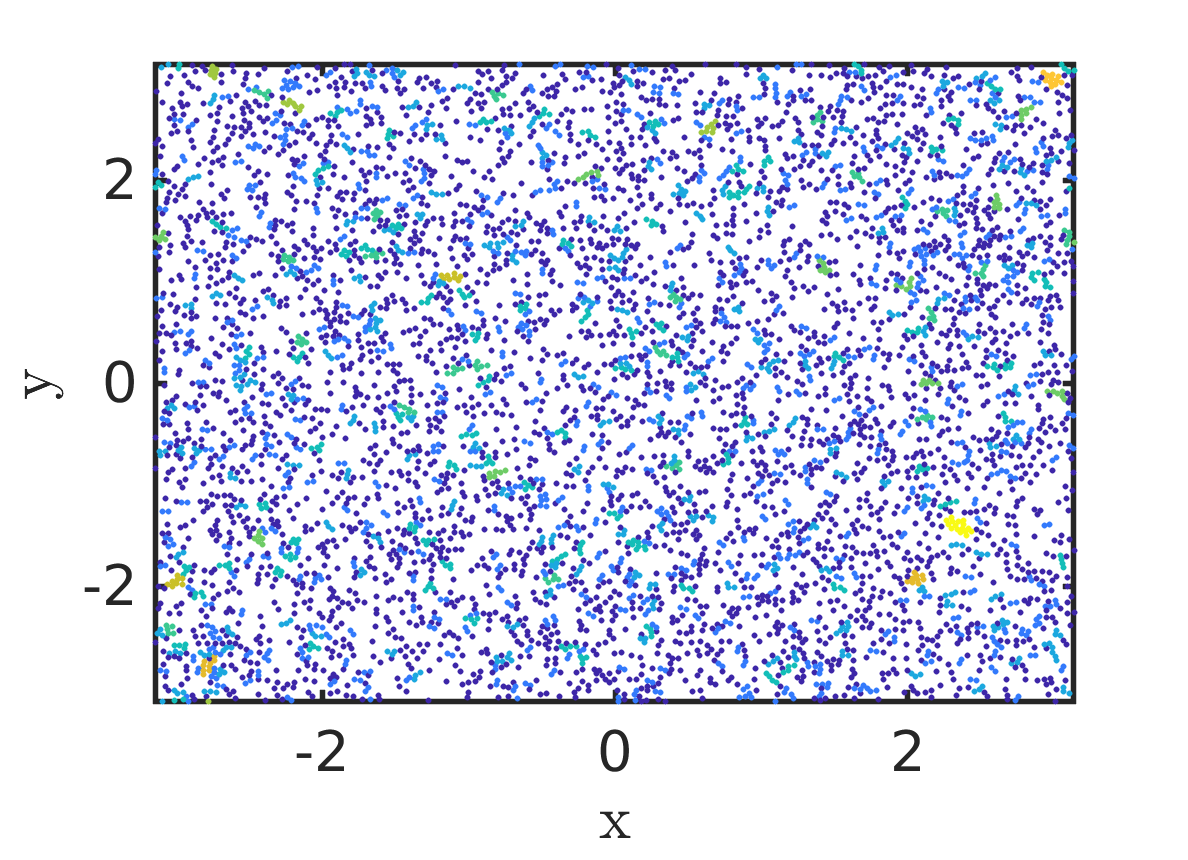}
    \put(-42,125){\colorbox{white}{\textcolor{black}{\large \bf (a)}}}
    \includegraphics[width=.3\linewidth,height=.3\linewidth]{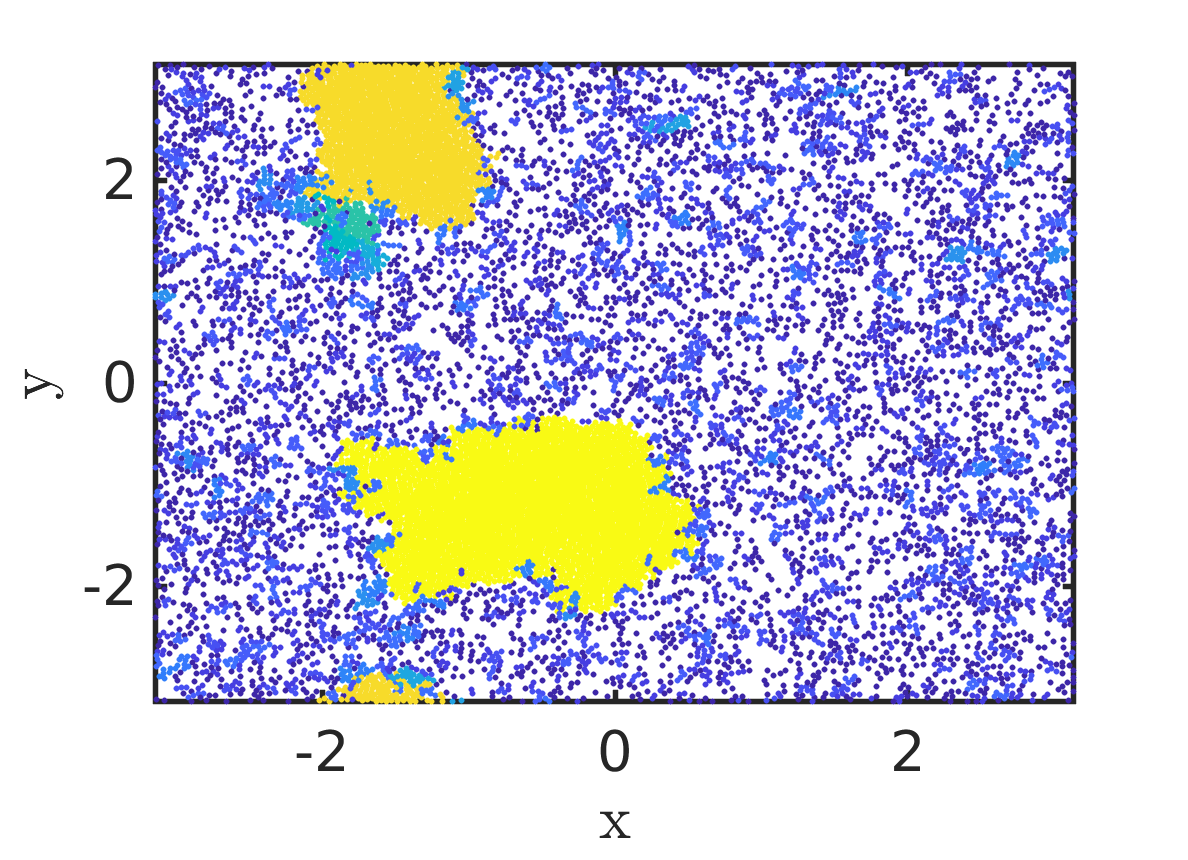}
    \put(-42,125){\colorbox{white}{\textcolor{black}{\large \bf (b)}}}
    \includegraphics[width=.3\linewidth,height=.3\linewidth]{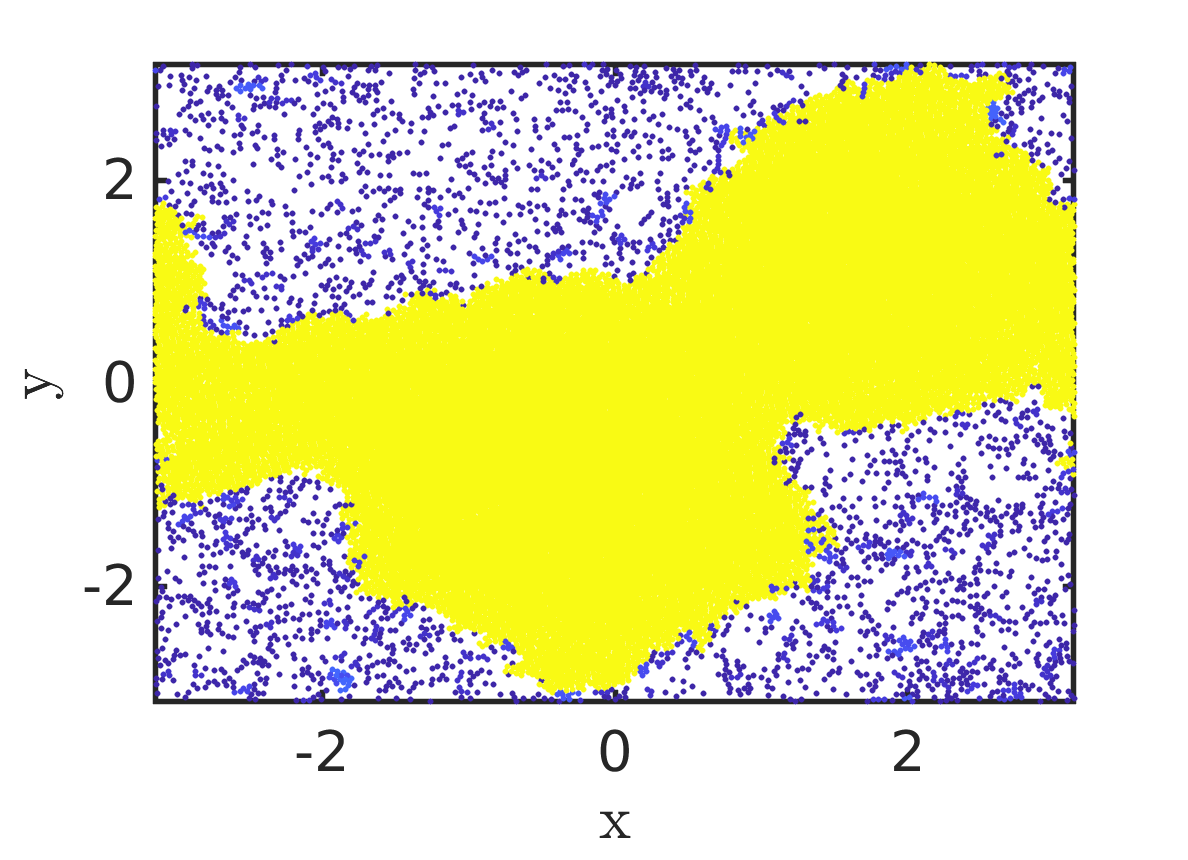} 
    \put(-42,125){\colorbox{white}{\textcolor{black}{\large \bf (c)}}}
    \\
    \includegraphics[width=.3\linewidth,height=.3\linewidth]{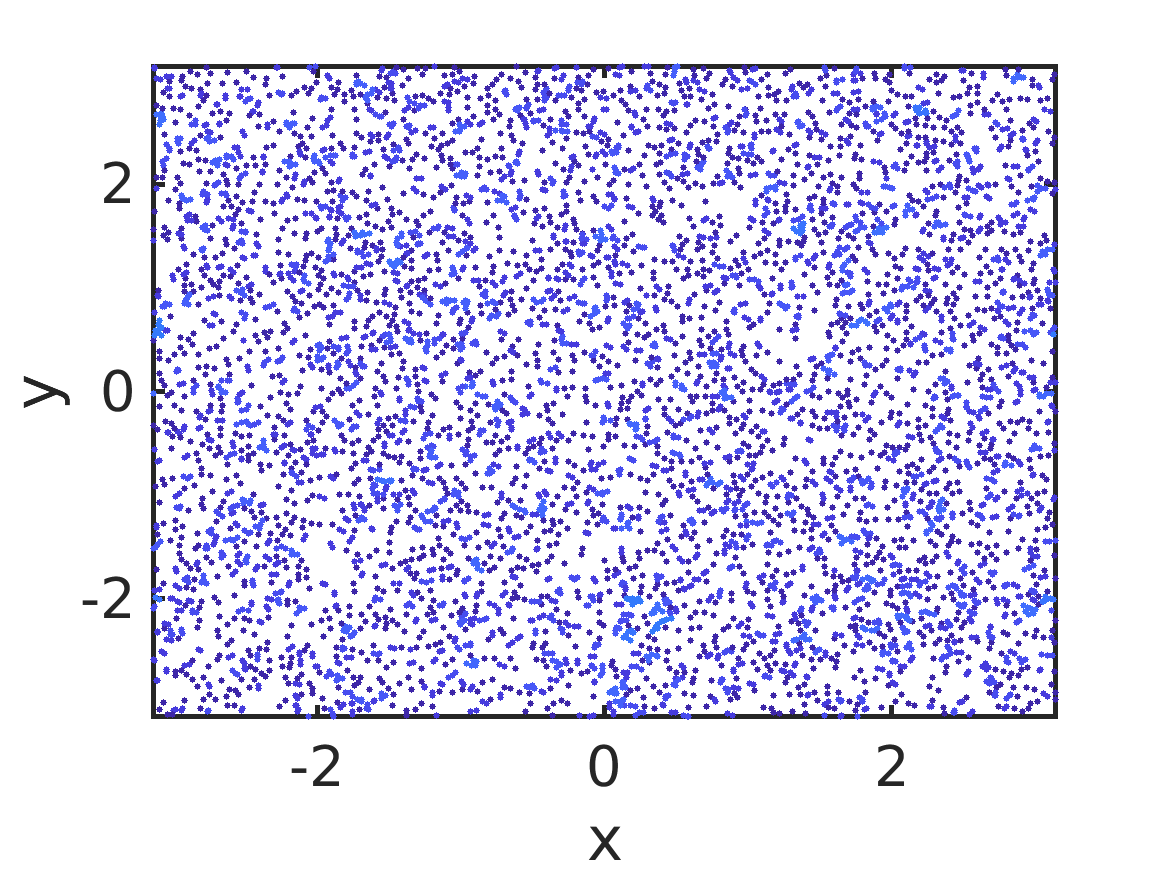}
    \put(-42,125){\colorbox{white}{\textcolor{black}{\large \bf (d)}}}
    \includegraphics[width=.3\linewidth,height=.3\linewidth]{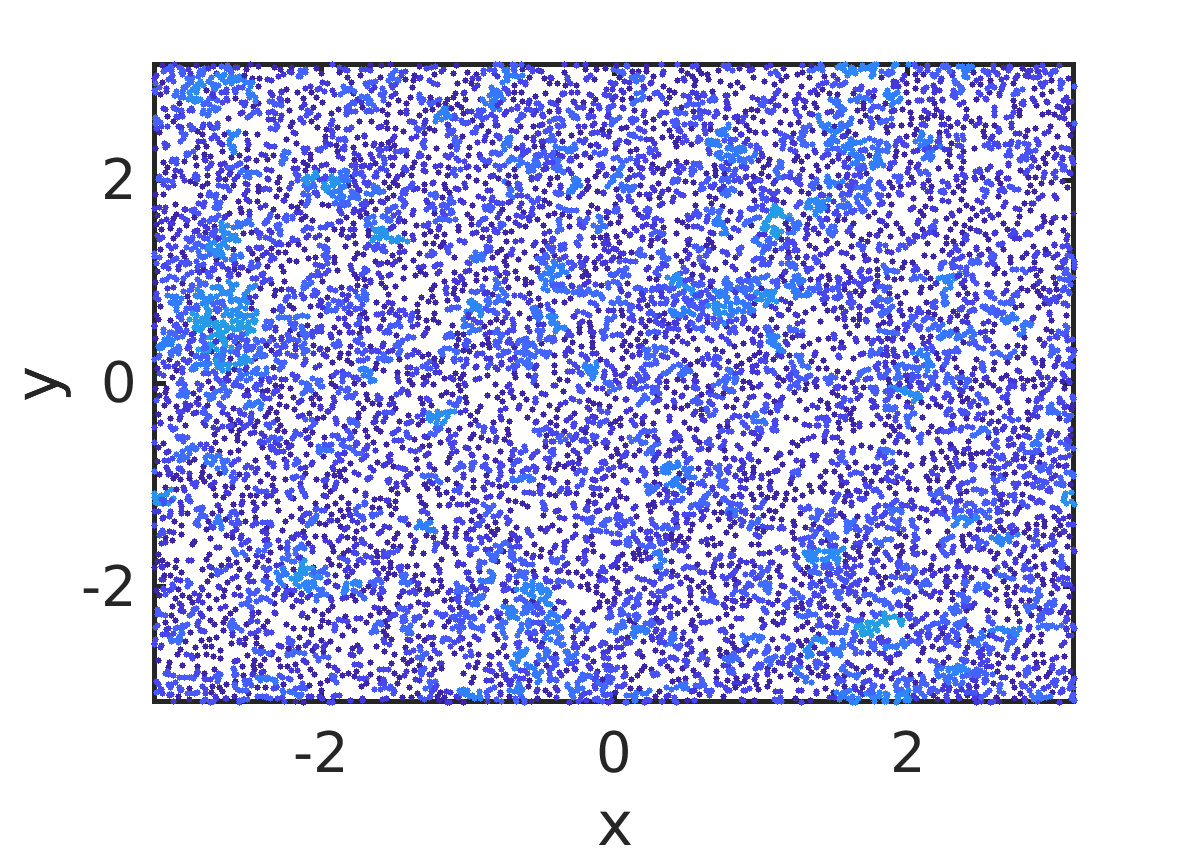}
    \put(-42,125){\colorbox{white}{\textcolor{black}{\large \bf (e)}}}
    \includegraphics[width=.3\linewidth,height=.3\linewidth]{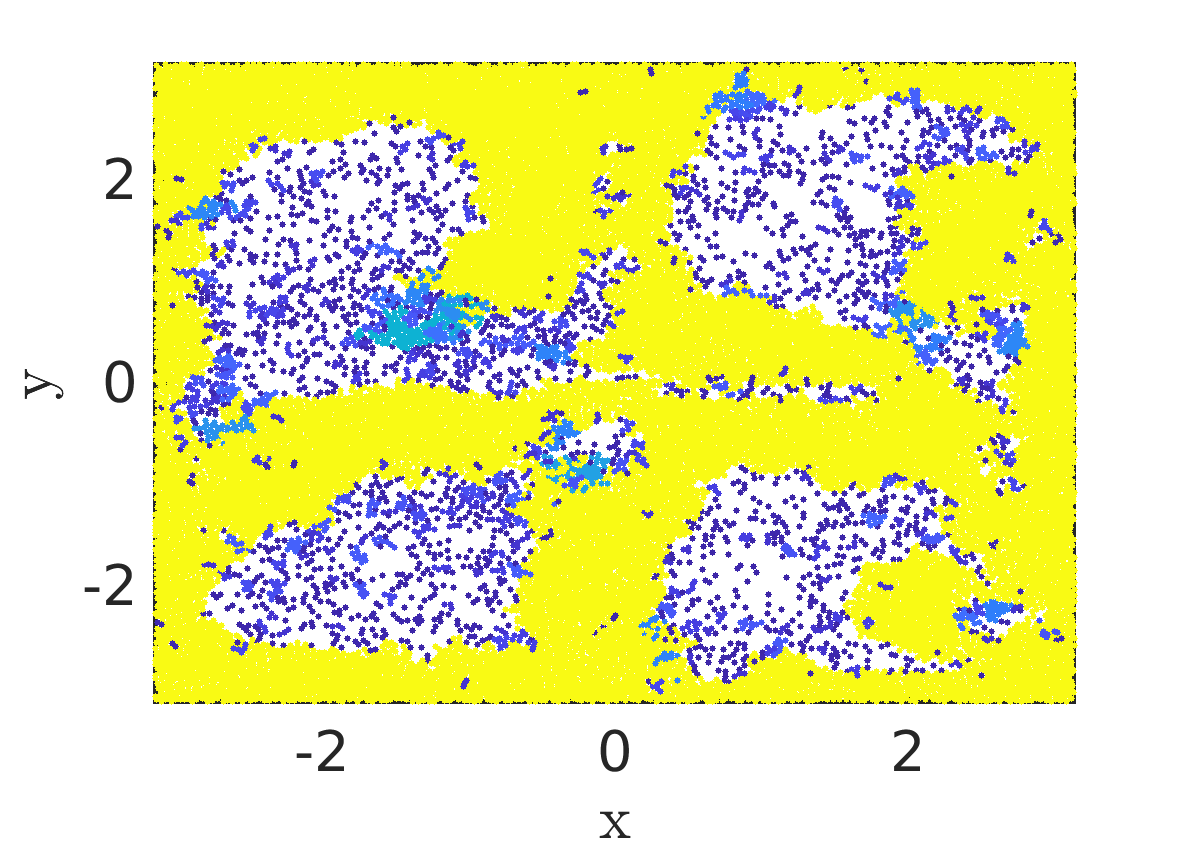}
    \put(-42,125){\colorbox{white}{\textcolor{black}{\large \bf (f)}}}
    \\
    \includegraphics[width=.8\linewidth,height=.08\linewidth]{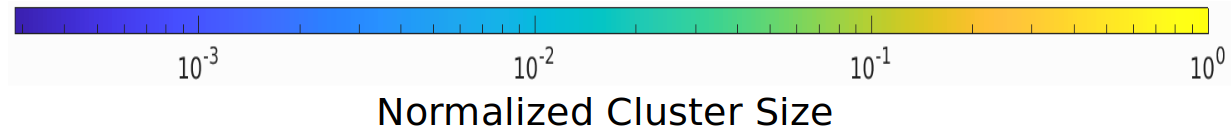}
    \caption{\label{fig:snap_plot_mf} \justifying{Representative snapshots of ABPs without (top panel) and with (bottom panel) imposed flow. The left, center, and right panels correspond to $\phi = 0.2, 0.4$, and $0.7$, respectively. The color bar represents the normalized cluster size.}}
\end{figure*}
\subsection{Flow-Induced Phase Separation}
\label{section3}
In Fig.~\ref{fig:snap_plot_mf}, we present the representative snapshots of particle configurations without (top panel) and with (bottom panel) background flow for three different packing fractions $\phi=0.2$ (left), $\phi=0.4$ (center), and $\phi=0.7$ (right). For low packing fractions ($\phi = 0.2$), no phase separation is observed in either case; active Brownian particles (ABPs) remain homogeneously distributed throughout the domain (Figs.~\ref{fig:snap_plot_mf}a,d). At intermediate packing fraction ($\phi = 0.4$), phase separation emerges in the absence of flow (Fig.~\ref{fig:snap_plot_mf}b), whereas under imposed flow the particles remain uniformly distributed (Fig.~\ref{fig:snap_plot_mf}e).

At high packing fraction ($\phi = 0.7$), both systems exhibit phase separation, though their morphologies differ qualitatively. Without flow, we observe standard MIPS, where dense clusters form spontaneously and diffuse throughout the domain (Fig.~\ref{fig:snap_plot_mf}c). In contrast, under imposed flow, the dense phase becomes localized in the strain-dominated regions, forming a characteristic four-lobe structure corresponding to the vortex cores of the four-roll mill flow (Fig.~\ref{fig:snap_plot_mf}f). A systematic variation of $\phi$ from $\rm \phi \in [0.1, 0.9]$ is shown in Appendix Fig.~\ref{fig:snap_plot}, demonstrating that even at very high packing fractions ($\phi = 0.9$), the four distinct low-density regions persist, coinciding with the four vortex cores of the background flow.

In this system, $\rm \mathcal{O}(v_p) = \mathcal{O}(U_0)$, while the flow field is spatially varying. Consequently, in regions where $\rm |\mathbf{u}| > v_p$, particle motion is primarily governed by the external flow, which typically corresponds to regions of low vorticity, causing particles to follow the streamlines. In contrast, in regions where $\rm |\mathbf{u}| < v_p$, the dynamics are dominated by the self-propulsion velocity of the particles. These regions are associated with larger vorticity, which strongly influences the particle orientation and direction of motion.
At low packing fractions, particles have sufficient free space to explore the domain, resulting in a nearly homogeneous spatial distribution (Fig.~\ref{fig:snap_plot_mf}(d) and ~\ref{fig:snap_plot_mf}(e)). However, at high packing fractions, interparticle interactions significantly affect the dynamics. Starting from a homogeneous configuration, particles in the strain-dominated regions largely follow the background flow, whereas particles within vortices experience strong vorticity, enabling them to escape the vortices and accumulate in the strain-dominated regions. This process ultimately leads to FIPS (Fig.~\ref{fig:snap_plot_mf}(f)). 
\begin{figure*}[!htpb]
    \centering
    \includegraphics[width=1.0\linewidth]{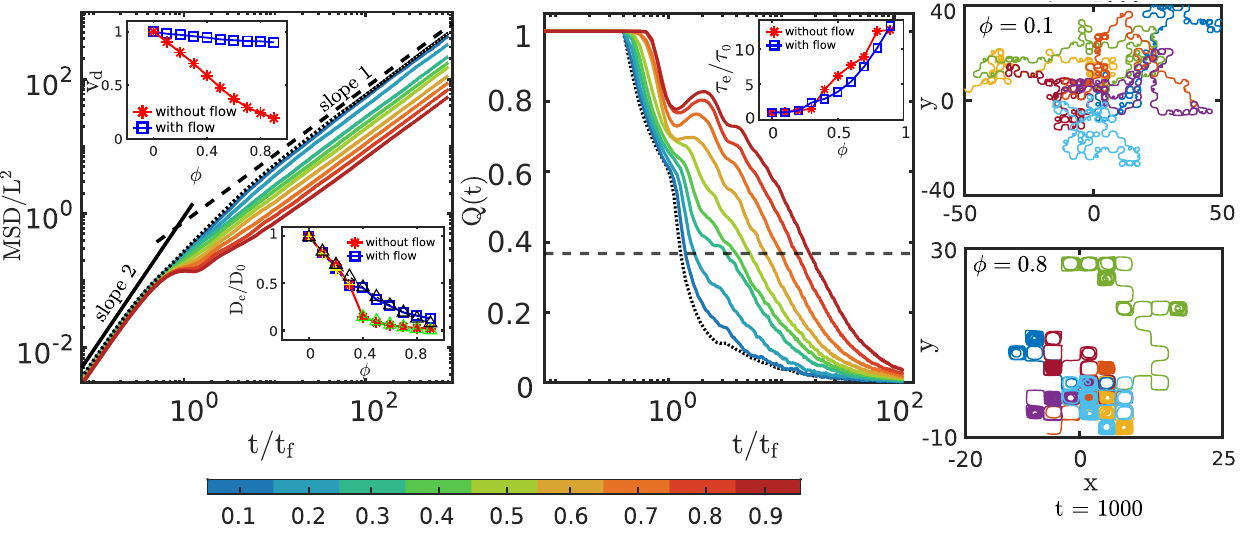}
    \put(-465,68){\textcolor{black}{\normalsize \bf (a)}}
     \put(-285,68){\textcolor{black}{\normalsize \bf (b)}}
     \put(-110,145){\textcolor{black}{\normalsize \bf (c)}}
     \put(-110,47){\textcolor{black}{\normalsize \bf (d)}}
    \caption{\justifying Dynamical behavior of ABPs under background flow.
(a) MSD normalized by the domain size as a function of time normalized by the flow timescale ($\rm t_f$) for packing fractions $\rm \phi=0.1$–$0.8$. A ballistic–diffusive crossover is observed, with an intermediate plateau at large $\rm \phi$ due to particle trapping in strain-dominated regions and near vortex peripheries, away from the cores, as illustrated by the trajectories in (d). Insets: Normalized drift velocity (top-left) and effective diffusivity (bottom-right) for cases with (squares) and without (asterisks) flow, obtained from MSD fits as functions of $\rm \phi$.
(b) Overlap function $\rm Q(t)$ versus time showing delayed relaxation at high packing fractions and an intermediate modulation associated with transient trapping in strain-dominated regions and vortex peripheries. The black dotted line in (a) and (b) corresponds to non-interacting ABPs in background flow. Inset: Normalized relaxation time versus $\rm \phi$ for FIPS and MIPS defined as $\rm Q(t) = 1/e$. While both increase exponentially with $\rm \phi$, MIPS shows a pronounced rise near $\rm \phi_c=0.3$. The relaxation time is normalized by the value in the non-interacting case.
(c) and (d) Particle trajectories at $\rm \phi=0.1$ and $0.8$, respectively, at simulation time $\rm t=1000$.}
    \label{fig:msd}
\end{figure*}
\vspace{-0.4mm}
\subsection{Mean Square Displacement (MSD)}
The mean-square displacement (MSD) of ABPs without interactions follows the well-known result \cite{Fily2012,howse2007self}:
\begin{equation}\label{eq:MSDful}
\rm \langle[\Delta \boldsymbol{r}(t)]^2\rangle = 4 D_{0} \left[t + \frac{1}{\nu_r}(e^{-\nu_r t}-1) \right],
\end{equation}
where $\rm D_0 = v_p^2 / (2 \nu_r)$ is the absolute diffusivity and $\rm v_p$ is the absolute self-propulsion speed \cite{romanczuk2012active}. When interactions are introduced, overall MSD decreases with increasing packing fraction $\phi$~\cite{Fily2012}.
       
Fig.~\ref{fig:msd}(a) shows the mean-square displacement (MSD), normalized by the domain size, for ABPs in the presence of background flow at different packing fractions $\phi \in [0.1, 0.9]$. For low to intermediate packing fractions $\rm \phi \le 0.5$, the system exhibits a characteristic ballistic-to-diffusive crossover approximately at fluid time scale $\rm t_f = \rm 2\pi/\rm \omega_{rms}$. 
This can be understood by examining the trajectories shown in Fig.~\ref{fig:msd}(c).
At short times $\rm t \ll t_f$, the particles move along nearly persistent paths, leading to ballistic scaling $\rm \langle[\Delta \boldsymbol{r}(t)]^2\rangle \sim t^2$. 
As $\rm t$ approaches the fluid time scale $\rm t_f$, the trajectories curve and lose directional correlation. For $\rm t \gtrsim t_f$, the motion becomes effectively randomized, resulting in diffusive behavior $\rm \langle[\Delta \boldsymbol{r}(t)]^2\rangle \sim t$. As the system approaches the FIPS regime in the high packing fraction limit ($\rm \phi \geq 0.6$), the MSD retains the overall ballistic-to-diffusive transition but develops an additional intermediate ``plateau" region~\cite{prajapati2025effect}. In Fig.~\ref{fig:msd}(d), at short times $\rm t \ll t_f$, particles exhibit nearly persistent motion, resulting in ballistic scaling. As $\rm t$ approaches the fluid time scale $\rm t_f$, particles become temporarily trapped between strain-dominated regions and near vortex peripheries, away from the vertex cores due to crowding, leading to a plateau in the MSD. Eventually, they escape these cages and exhibit diffusive motion for $\rm t \gtrsim t_f$. 
In Fig.~\ref{fig:msd}(a), the value $\rm \mathrm{MSD/L^2} \big|_{\mathrm{t = t_f}} \sim \rm 0.1-0.2$ is consistent with the expected vortex size, which is approximately $\rm L/4-L/2$. To highlight the role of interactions, we additionally plotted the MSD of non-interacting ABPs in a background flow, indicated by the black dotted line in Fig.~\ref{fig:msd}(a), which shows good agreement at low packing fractions.


\vspace{3mm}
\textit{Drift Velocity $\rm (v_d)$ and Diffusivity $\rm (D_e)$:} 
From the MSD curves, we determine both the drift velocity, $\mathrm{v_d}$, and the effective diffusivity, $\mathrm{D_e}$. The drift velocity of the ABPs is normalized by a characteristic velocity scale obtained from the non-interacting case. In the absence of flow, this scale corresponds to the self-propulsion speed $\vp$, whereas in the presence of flow it is given by the characteristic velocity, i.e., root-mean-square velocity, $\mathrm{U_{rms}}$, of the background flow field $\rm \mathbf{u}(\mathbf{r},t)$.
In the absence of background flow, the normalized drift velocity $\rm \overline{v}_d$ decreases with packing fraction $\phi$. The effective propulsion speed follows $\rm v_d(\phi)=v_p(1-\lambda\phi)$ with $\lambda=0.9$, shown in the upper inset of Fig.~\ref{fig:msd} (red asterisk)~\cite{Fily2012}.

In the presence of the flow, the $\rm \overline{v}_d$ is nearly constant which indicate that the normalized drift velocity ($\rm \overline{v}_d$) is primarily driven by the background flow, which corresponds to $\mathrm{ U_{rms}} = \mathrm{ U_{0}}/\sqrt 2$, as shown in the upper inset of Fig.~\ref{fig:msd}(a) in blue square, for all values of $\phi$.
 This suggests that self-propulsion and flow together maintain a sustained translation irrespective of crowding, whereas for MIPS, there is a strong dependence on crowding.
 
Similarly, the extracted effective diffusivity is normalized by the absolute diffusivity of the corresponding non-interacting system. For the MIPS case, this diffusivity is given by $\rm D_0 = v_p^2/2\nu_r$~\cite{Fily2012}. In contrast, for the FIPS case we obtain $\rm D_0 \sim 0.5$. Determining the exact analytical form of the absolute diffusivity in the presence of flow is beyond the scope of the present work and will be addressed in future studies.

As shown in the lower inset of Fig.~\ref{fig:msd}(a), the normalized effective diffusivity $\rm D_e/D_0$ decreases with increasing packing fraction $\rm \phi$. In the MIPS case, this decrease exhibits a sharp transition near the critical packing fraction $\rm \phi_c \sim 0.3$, consistent with earlier reports~\cite{PhysRevResearch.7.013153}. In contrast, for FIPS, $\rm D_e/D_0$ decreases smoothly with $\rm \phi$, indicating a continuous transition.
For MIPS, the data are well described by $\rm D_e = D_0(1-\lambda_1 \phi^2)$ for $\rm \phi < 0.3$ (yellow plus markers), with $\rm \lambda_1 = 1$. For $\rm \phi > 0.3$, the behavior is better captured by $\rm D_e = D_0 \lambda_2 (1-\lambda_1 \phi)^2$, with $\rm \lambda_2 = 0.4$ (green triangles), as reported by Nayak \textit{et al.}~\cite{PhysRevResearch.7.013153}. In the FIPS case, the diffusivity follows a different scaling, $\rm D_e = D_0(1-\lambda_3 \phi)$, with $\rm \lambda_3 = 0.8$. 
This indicates that particles move less freely at higher densities because they are more crowded and interact more frequently, slowing their overall motion.
\subsection{Overlap Function}
Fig~\ref{fig:msd}(b) is the typical curve of time-dependent self overlap function $\rm Q(t)$ for different $\rm \phi$ is defined as
\begin{equation}\label{eq:overlap}
    \mathrm{Q(t) = \frac{1}{N}\langle \sum_{j = 1}^{N}} w(|\mathrm{\mathbf{r}_j(t) - \mathbf{r}_j(0)}|)\rangle,
\end{equation}
where $\rm N$ denotes the number of particles and the weight function $w(\mathrm{x})$ is defined as $w(\mathrm{x})=1$ for $\rm x < \pi$ (i.e. within the vortex size) and $w(\mathrm{x})=0$ otherwise~\cite{mandal2020extreme},\cite{karmakar2011exposingstaticscaleglass}.

Fig~\ref{fig:msd}(b) shows $\rm Q(t)$ exhibits an approximately exponential decay with some modulations depends on $\rm \phi$. At low packing fractions, particles escape their cages before the flow time scale $\rm t_f$, indicating that the decorrelation dynamics are governed by both the flow and the intrinsic activity of the particles. The self-propulsion enhances particle motion and consequently reduces the cage escape time. Around $\rm t = t_f$, a small modulation in the $\rm Q(t)$ curve suggests that the flow tends to recage the particles due to its periodic nature; however, the persistent activity prevents sustained recaging.

After the critical packing fraction $\rm \phi_c \gtrsim 0.5$ for FIPS, particle interactions become dominant, which delays the cage-out process. Near $\rm t = t_f$, recaging is enhanced due to crowding between vortices and strain-dominated regions. The clusters as a whole become trapped in and around the strain-dominated regions, as indicated by the pronounced modulation in the $\rm Q(t)$ curve. Eventually, particles escape the cage due to their self-propelled velocity.

The $\rm Q(t)$ plot shows stronger caging and recaging at high packing fractions, indicating that particles remain trapped for longer times before escaping, consistent with the plateau observed in the MSD. To highlight the role of interactions, we also plot $\rm Q(t)$ for non-interacting ABPs in the same flow field, which shows good agreement with the interacting case at low packing fractions.

\vspace{3mm}
\textit{Relaxation time:}
The relaxation time, $\rm \tau_e$, is defined as the time at which the correlation function satisfies $\rm Q(t)=1/e$, where $\rm e$ is the base of the natural logarithm~\cite{karmakar2011exposingstaticscaleglass}. The inset of Fig.~\ref{fig:msd}(b) shows the normalized relaxation time, $\rm \tau_e/\tau_0$, as a function of the packing fraction $\rm \phi$ for both without and with background flow cases, where $\rm \tau_0$ denotes the relaxation time in the non-interacting case.

For both cases, the relaxation time is growing exponentially as $\rm \phi$ increases which means basic slowdown mechanism is interaction-driven in both cases. The system without flow exhibits a relatively sharp increase around $\rm \phi \approx 0.3$, indicating a sudden onset of clustering due to activity-driven phase separation. In contrast, the FIPS case shows a smoother variation of the relaxation time, suggesting that the background flow continuously redistributes particles and leads to a more gradual buildup of caging and clustering. 
\begin{figure*}[!htb]
    \centering
    \includegraphics[width=1.0\linewidth]{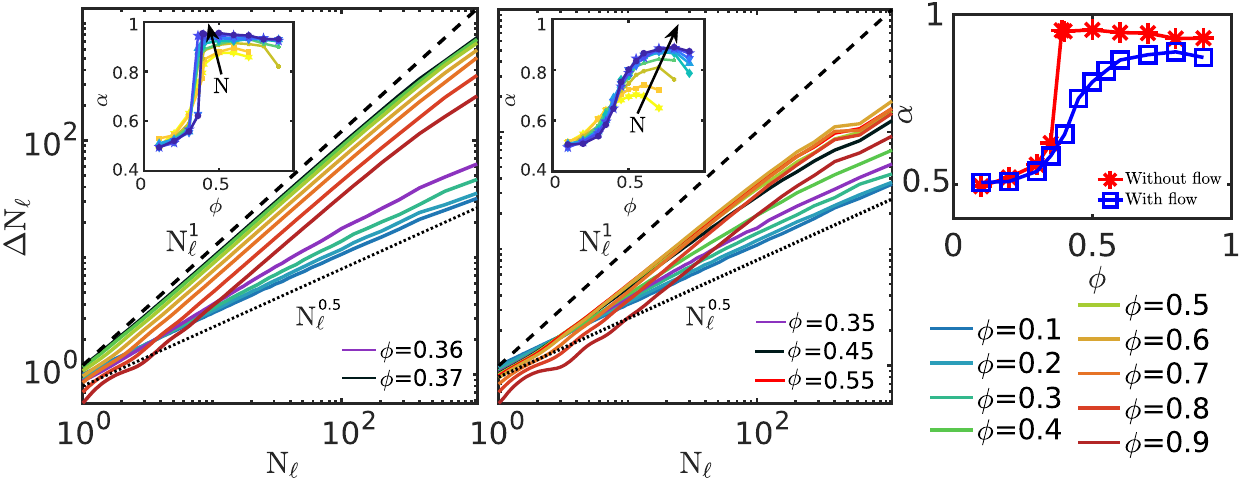}
    \put(-460,35){\textcolor{black}{\normalsize \bf (a)}}
     \put(-288,35){\textcolor{black}{\normalsize \bf (b)}}
     \put(-110,112){\textcolor{black}{\normalsize \bf (c)}}
    \caption{\justifying Number fluctuations in ABPs varying with packing fraction for (a) system without background flow shows giant number fluctuations for $\phi \ge 0.36$ and (b) system with flow shows giant number fluctuations for $\phi \gtrsim 0.5$.
Insets in (a) and (b) show the scaling exponent $\alpha$ of the number fluctuations as a function of packing fraction \(\phi\) for different system sizes, \(\rm N = 880\)–\(10^{4}\), providing a good statistics for large number of particles for both without and with flow cases. (c) A direct comparison between without and with flow cases exhibits a sharp and smooth transition, respectively.}
    \label{fig:std_dev}
\end{figure*}
\subsection{Number Fluctuations}
\label{section5}

To characterize clustering, we analyze the number fluctuations $\rm \Delta N_\ell$ within the square subsystems of area $\rm \ell^2$, and study their scaling with the mean particle number $\rm N_\ell$. Specifically, we examine the scaling behavior $\rm \Delta N_\ell \sim N_\ell^\alpha$, where the exponent $\rm \alpha$ captures the nature of fluctuations. The value of the exponent $\rm \alpha = 0.5$ for a homogeneous distribution. In contrast, especially those undergoing phase separation, can exhibit giant number fluctuations characterized by $\rm \alpha \to 1$, signaling strong correlations and spontaneous clustering \cite{Fily2012, henkes2011active, Narayan2007, kuroda2023anomalous, prajapati2025effect}.

Fig.~\ref{fig:std_dev}(a) shows the giant number fluctuations of ABPs in the absence of any background flow for packing fractions ranging from $\rm \phi = 0.1$ to $\rm 0.9$. At low densities ($\rm \phi < 0.4$), we observe $\rm \alpha \approx 0.5$ (see the inset of Fig.~\ref{fig:std_dev}(a)), consistent with homogeneous distribution, which implies that there is no phase separation. As the system enters the MIPS regime at higher densities ($\rm \phi \geq 0.4$), the exponent increases towards $\rm \alpha \approx 1$, indicating the emergence of large-scale density inhomogeneities and clustering driven by persistent self-propulsion \cite{Fily2012}.

In the presence of structured background flow, for $\rm \phi \leq 0.4$, the system exhibits equilibrium-like number fluctuations, $\rm \alpha \approx 0.5$ clearly indicated in the inset of  Fig.~\ref{fig:std_dev}(b). However, beyond $\rm \phi > 0.5$, the exponent increases toward unity, marking the onset of FIPS, where the combined effects of activity, crowding, and external flow drive the formation of dense, dynamic clusters \cite{prajapati2025effect}. Here, The inset of Fig.~\ref{fig:std_dev}(a) and ~\ref{fig:std_dev}(b) shows the scaling exponent as a function of $\phi$ for different system sizes ranging from $880$ to $10^4$. For sufficiently large particle numbers, giant number fluctuations are evident, providing reliable statistics for both MIPS and FIPS case.

Furthermore, MIPS emerges at relatively lower packing fractions $\rm \phi$ compared to the FIPS scenario. The background flow enhances particle mobility, which suppresses or delays clustering. As shown in Fig.~\ref{fig:std_dev}(c), the MIPS case exhibits systematically larger values of $\rm \alpha$ than the FIPS case, indicating stronger number fluctuations and more pronounced density inhomogeneities in the absence of flow.
The onset of giant number fluctuations is significantly sharper in the MIPS case, whereas in FIPS the increase of the fluctuation exponent occurs more gradually, as shown in Fig.~\ref{fig:std_dev}(c). This qualitative difference suggests a discontinuous-like transition in MIPS and a more continuous-like transition in FIPS. However, this inference is based solely on the behavior of the number-fluctuation exponent; a rigorous characterization of the transition and its order lies beyond the scope of the present work and will be addressed in a future study.

\begin{figure*}[!htb]
    \centering
    \includegraphics[width=1.0\linewidth]{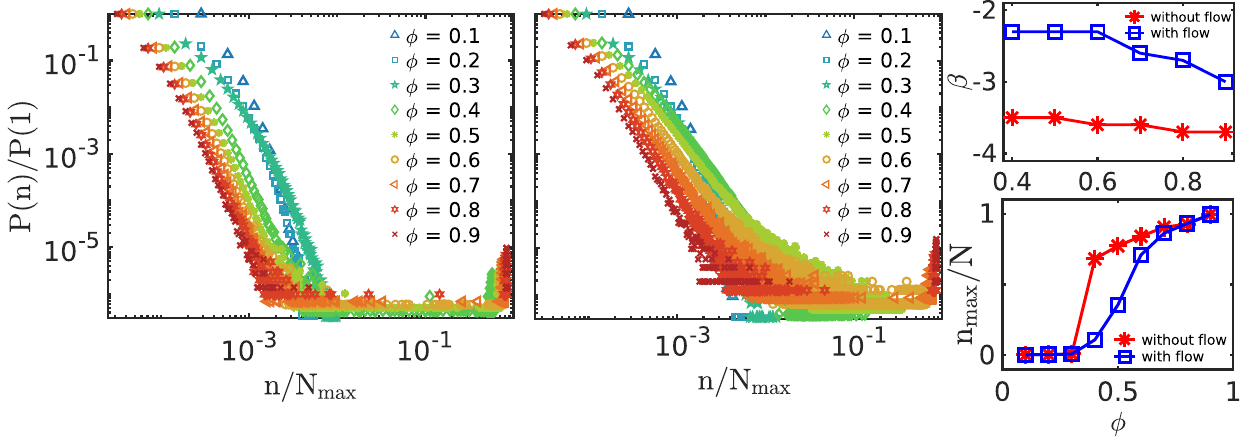}
    \put(-460,55){\textcolor{black}{\normalsize \bf (a)}}
    \put(-285,55){\textcolor{black}{\normalsize \bf (b)}}
    \put(-95,120){\textcolor{black}{\normalsize \bf (c)}}
    \put(-95,90){\textcolor{black}{\normalsize \bf (d)}}
    \caption{\justifying Normalized cluster size distribution $\rm P(n)/P(1)$ as a function of ratio of cluster size to maximum number of particle at corresponding packing fractions $\rm n/N_{max}$ for ABPs with size $\rm a = 0.02$ at various packing fractions $\rm \phi \in [0.1, 0.9]$: (a) without background flow exhibiting MIPS behavior, and (b) with background flow exhibiting FIPS behavior. (c) Exponent of the cluster size distribution (CSD) and (d) normalized maximum cluster size shown as functions of packing fraction $\phi$, comparing systems in the absence and presence of background flow.} 
    \label{fig:CSD_compare}
\end{figure*}
\subsection{Cluster Size Distribution}

We analyze the cluster size distribution (CSD) to characterize the nature of clustering in both the absence and presence of background flow. In Fig.~\ref{fig:CSD_compare}, we plot the normalized cluster size distribution, $\rm P(n)/P(1)$, as a function of cluster size $\rm n$, where $\rm P(n)$ is the probability of finding a cluster of size $\rm n$ composed of ABPs at different packing fractions $\rm \phi \in [0.1, 0.9]$. The normalized CSD is well described by a power law with an exponential cutoff of the form ~\cite{prajapati2025effect, dolai2018phase}:
\begin{equation}
    \rm P(n)/P(1) \simeq \frac{1}{n^\beta} \exp\left(-\frac{n}{n_0}\right),
\end{equation}
where $\rm \beta$ is the power-law exponent and $\rm n_0$ is the cutoff cluster size.

In the absence of background flow (Fig.~\ref{fig:CSD_compare}(a)), the CSD exhibits a distinct behavior depending on the packing fraction. For $\rm \phi < 0.4$, corresponding to a homogeneous phase, the CSD is unimodal, indicating the absence of large-scale clustering. For higher packing fractions $\rm \phi \ge 0.4$, where MIPS occurs, the CSD shows bimodal distribution, signaling the coexistence of large dense clusters with a gas-like phase \cite{Fily2012, dolai2018phase, sanoria2022percolation}.

When we introduced background flow (Fig.~\ref{fig:CSD_compare}(b)), the transition from unimodal to bimodal CSD shifts to higher packing fractions, occurring beyond $\rm \phi > 0.5$. This shift reflects the suppression or delay of phase separation due to presence of flow. Furthermore, the power-law exponent $\rm \beta$ exhibits a dependence on $\rm \phi$, suggesting an interplay between flow and clustering dynamics. Such behavior is consistent with observations in FIPS systems \cite{prajapati2025effect}. 

We observed that the exponent $\beta$ for $\phi \ge 0.4$, as shown in Fig.~\ref{fig:CSD_compare}(c), exhibits a clear difference between MIPS (red asterisk) and FIPS (blue squares). The exponent $\beta$ is consistently higher in the presence of background flow, indicating the formation of smaller clusters due to the flow. In addition, we also observed the normalized maximum cluster size $\rm n_{max}/N$ across all values of $\phi$, where $\rm n_{max}$ is the largest cluster size (containing $\rm n$ particles), and $\rm N$ is the total number of particles corresponding to given packing fractions. As shown in Fig.~\ref{fig:CSD_compare}(b), there is clear evidence that the presence of flow suppresses cluster formation.
\subsection{PDF of Okubo-Weiss parameter}

To further validate the clustering behavior of particles in the flow, we compute the probability distribution function (PDF) of the Okubo--Weiss parameter, \(\Lambda = (\omega^2 -\sigma^2) /8 \) \cite{prajapati2025effect}, following the approach in Refs.~\cite{picardo2018preferential, picardo2020dynamics}. This parameter distinguishes between vortical (\(\Lambda > 0\)) and strain-dominated (\(\Lambda < 0\)) regions of the flow.

\begin{figure}[!htp]
    \centering
    \includegraphics[width=1\linewidth]{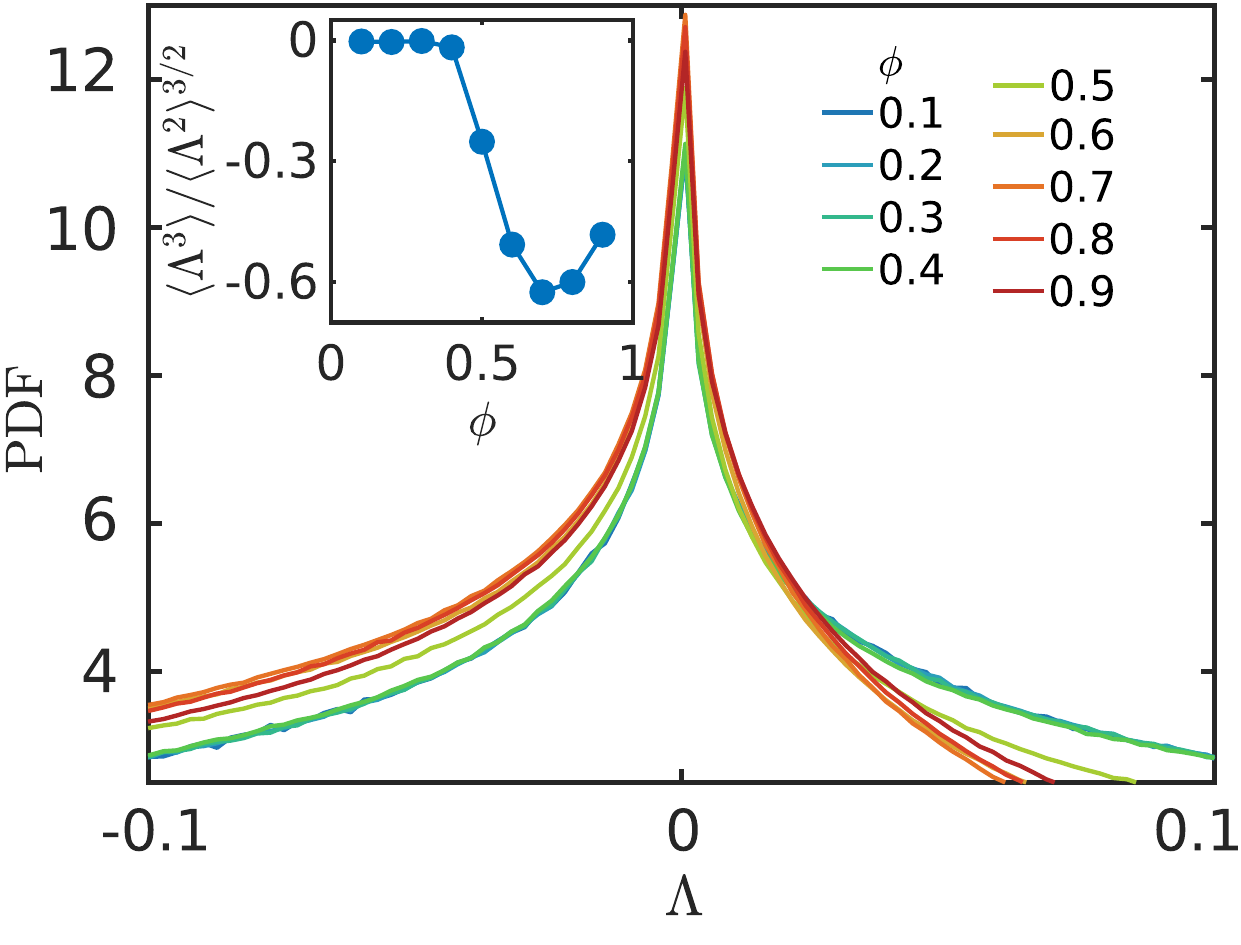}
    \caption{\justifying PDF of the Okubo–Weiss parameter $\rm \Lambda$ for different $\rm \phi \in [0.1, 0.9]$ values. The emergence of asymmetry in the distribution with increasing $\phi$ indicates growing localization in strain-dominated regions. Inset: Plot of skewness ($\rm \frac{\langle \Lambda^3 \rangle}{ \langle \Lambda^2) \rangle^{3/2}}$) versus $\phi$.}
    \label{fig:okubo_diff_nd}
\end{figure}

Fig.~\ref{fig:okubo_diff_nd} presents the PDF of \(\Lambda\) sampled along particle trajectories for varying particle packing fractions \(\phi\). At low packing fractions (\(\phi \lesssim 0.4\)), the PDF is nearly symmetric about \(\Lambda = 0\), indicating a relatively uniform distribution of particles between vortical and strain-dominated regions. This suggests minimal preferential sampling by the particles at low packing fraction.
However, as the packing fraction increases (\(\phi \gtrsim 0.5\)), the PDF becomes noticeably asymmetric, exhibiting a pronounced tail for negative values of \(\Lambda\). This negative skewness indicates an increased likelihood of particles residing in strain-dominated regions of the flow, as shown in inset of Fig. \ref{fig:okubo_diff_nd}. 

The inset of Fig.~\ref{fig:okubo_diff_nd} reveals an interesting trend: beyond $\phi = 0.7$, the magnitude of skewness shows a slight decrease, which can be attributed to increased particle crowding. As the packing fraction rises, the finite size of particles forces a larger fraction of them into the vortical regions of the flow, as also evident from the snapshots in Appendix Fig.~\ref{fig:snap_plot}. This redistribution reduces the overall skewness magnitude. Nevertheless, the skewness remains negative across all densities, indicating that most particles continue to occupy the strain-dominated regions of the flow.
\section{Conclusion and Discussion}
\label{section:conclusion}

In this study, we explored the collective dynamics of ABPs in a two-dimensional fluid environment under the influence of a steady background flow. By introducing a four-roll mill flow, we systematically investigated the structural and dynamical transitions in the system by varying the packing fraction $\rm \phi$ from $\rm 0.1$ to $\rm 0.9$.


Our results reveal a clear crossover from a homogeneous phase at low densities to a distinct FIPS regime at higher densities ($\rm \phi \gtrsim 0.5$). Unlike classical MIPS, which emerges due to self-trapping in crowded environments, FIPS arises from the interplay between particle motility, interparticle interactions, and external flow-induced vorticity. The emergence of persistent clusters in strain-dominated regions, stabilized by the background flow, characterizes this FIPS phase.

Through mean-square displacement (MSD) analysis, we observed a non-trivial crossover from ballistic to diffusive behavior with an intermediate plateau, particularly in the FIPS regime, indicating transient trapping. The drift velocity of the ABPs remained nearly constant across $\rm \phi$, mainly driven by the steady flow field, while the effective diffusivity decayed quadratically with $\rm \phi$, consistent with enhanced clustering and reduced mobility at higher densities.

The overlap function provides further insight into the origin of the plateau observed in the mean squared displacement. It indicates that periodic flow induces transient caging, which is temporarily disrupted by activity and followed by transient recaging due to crowding in strain-dominated regions. The relaxation time exhibits an exponential decay in both cases; however, its dependence on packing fraction differs between the MIPS and FIPS regimes. While particle interactions primarily drive the dynamical slowdown, the imposed flow redistributes particles in the system, leading to a more gradual development of caging and clustering.

Furthermore, number fluctuations exhibited a transition from normal ($\rm \Delta N_\ell \sim N_\ell^{1/2}$) to giant scaling ($\rm \Delta N_\ell \sim N_\ell$), signifying the onset of long-range density inhomogeneities in the FIPS phase. 
The cluster size distribution (CSD) shows that, in the absence of flow, MIPS emerges beyond a critical packing fraction (\(\phi \geq 0.4\)), characterized by a transition from a unimodal to a bimodal CSD. In contrast, background flow delays this transition, shifting the onset of phase separation to higher \(\phi\) values (\(\phi > 0.5\)). The probability distribution function (PDF) of the Okubo--Weiss parameter reveals that at low packing fractions, particles are uniformly distributed across strain- and vorticity-dominated regions. However, with increasing \(\phi\), the PDF becomes increasingly skewed toward strain-dominated regions, indicating preferential sampling and enhanced clustering in strain-dominated regions.

These findings enrich our understanding of active matter in complex environments, particularly in systems with external flow fields. The identification of FIPS opens new pathways for designing controllable clustering phenomena in synthetic microswimmer systems and may offer insights into biological transport and collective behavior in fluidic environments.


Our findings open up several avenues for future exploration. Incorporating thermal fluctuations would help assess the robustness of FIPS under realistic conditions, while extending the system to three dimensions may reveal more intricate clustering behavior. Allowing feedback from ABPs to the flow field, i.e., accounting for hydrodynamic interactions, would move the model closer to experimental systems such as bacterial suspensions. Exploring other flow topologies, such as shear or turbulent-like flows, could uncover new phases or transitions. Furthermore, studying anisotropic particles or alternative interaction potentials might broaden the observed collective behaviors. Finally, constructing effective theoretical models and comparing predictions with experimental realizations in microfluidic or colloidal systems could significantly advance our understanding of the flow-induced phase separation (FIPS).

\section{Acknowledgment}
SDP acknowledges Mustansir Barma for a fruitful discussion and suggestions. 
SDP acknowledges the receipt of University Grants Commission (UGC) 1608(CSIRNETJUNE2019), which supported this study.
AG acknowledges the funding from SERB-India (grant no. MTR/2022/000232, grant no. CRG/2023/007056-G); DST-India (grant no.DST/NSM/R\&D HPC Applications/2021/05 and grant no. SR/FST/PSI-309 215/2016). KS acknowledges the DST-INSPIRE Grant (No: DST/INSPIRE Fellowship/[IF230178]) for financial support.
\section{Appendix} \nonumber

\begin{figure*}[!htb]
    \centering
    \includegraphics[width=0.7\linewidth]{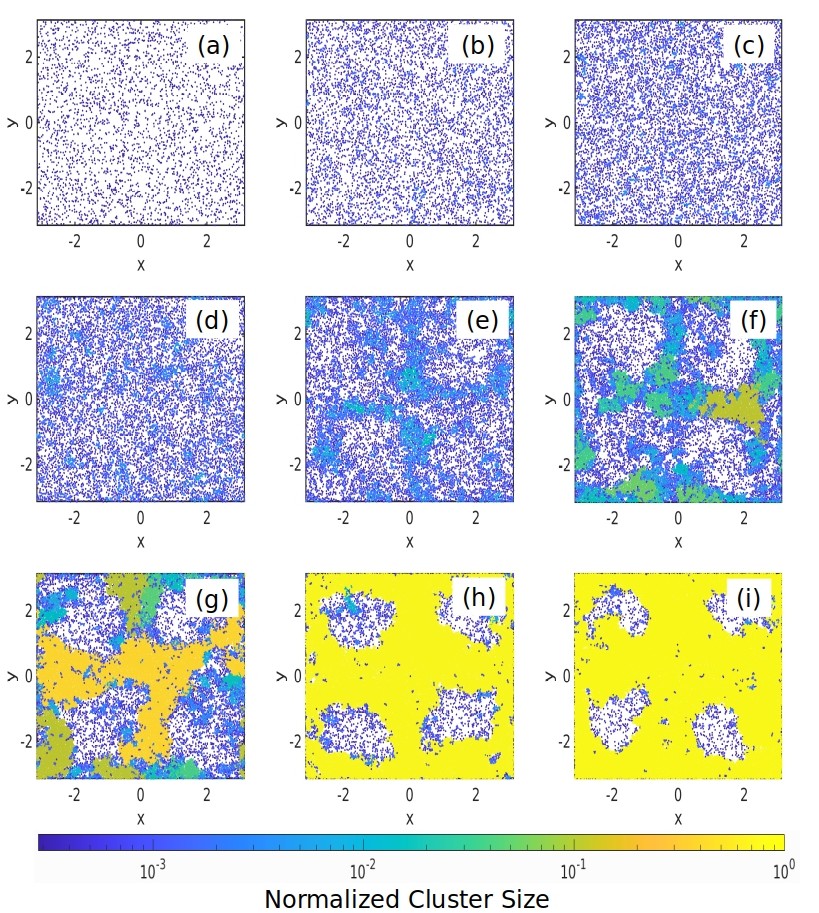}
    \caption{\label{fig:snap_plot} \justifying{Representative snapshots of ABPs for scaled velocity $\rm V = 0.4$ and particle size $\rm a = 0.02$, shown for various packing fractions $\phi \in [0.1, 0.9]$. The color scale represents the normalized cluster size, illustrating the transition from a homogeneous phase at low $\phi$ to a FIPS state at high $\phi$.}}
\end{figure*}

In this section, we present the spatial distribution of active Brownian
particles (ABPs) under the four-roll-mill background flow for various packing
fractions, $\phi$, ranging from 0.1 to 0.9 (Fig.~\ref{fig:snap_plot}). For low
packing fractions ($\phi \le 0.4$), the ABPs remain homogeneously distributed
throughout the domain. At $\phi = 0.5$, weak clustering begins to emerge near
the strain-dominated regions of the flow. For higher packing fractions ($\phi
\ge 0.6$), a clear flow-induced phase separation (FIPS) develops, characterized
by dense clusters anchored around the strain-dominated zones and four distinct
voids corresponding to the vortex cores of the four-roll-mill flow.

\newpage
\bibliography{FIPS24}

\end{document}